\documentclass[runningheads]{style/llncs}

\usepackage[T1]{fontenc}

\usepackage{orcidlink}
\renewcommand{\orcidID}[1]{\,\orcidlink{#1}}

\usepackage{graphicx}
\usepackage[list=true, font=small, labelfont=bf, 
labelformat=brace, position=top]{subcaption}

\usepackage{enumitem}

\usepackage{makecell}
\usepackage{tabularx}
\usepackage{booktabs}

\newcolumntype{C}{>{\centering\arraybackslash}X}

\usepackage{amsmath,amsfonts}
\usepackage{bm} % after amsmath

\usepackage{marvosym} 

\newcommand{\repourl}{https://github.com/Visual-Computing/sisap26-deglib}

\begin{document}
%
% If the paper title is too long for the running head, you can set an abbreviated paper title here
\title{Fast and Efficient Approximate Nearest Neighbor Search for High-Dimensional LLM Embeddings}
\titlerunning{Fast and Efficient ANNS for LLM Embeddings}

\author{
    Nico Hezel\inst{1}\orcidID{0000-0002-3957-4672}\,\textsuperscript{(\Letter)} \and
    Kai Uwe Barthel\inst{1,2}\orcidID{0000-0001-6309-572X} \and
    Bruno Schilling\inst{1}\orcidID{0009-0006-7021-7311} \and
    \\Konstantin Schall\inst{1,2}\orcidID{0000-0003-3548-0537} \and
    Andre Moelle\inst{2}\orcidID{0009-0009-1031-1949} \and
    Klaus Jung\inst{1}\orcidID{0000-0002-3600-6848}
}
%
% First names are abbreviated in the running head.
% If there are more than two authors, 'et al.' is used.
\authorrunning{N. Hezel et al.}

\institute{
    Visual Computing Group, HTW Berlin, Germany \url{www.visual-computing.com}\\
    \Letter\hspace{0.5em}\email{nico.hezel@htw-berlin.de}
    \and
    vviinn, Berlin, Germany
    \url{www.vviinn.com}
}

% --- ARXIV FOOTNOTE (NUR SEITE 1) ---
\makeatletter
\def\ps@arxivnote{%
  \def\@oddfoot{\reset@font\hfil\parbox{\textwidth}{\centering\color{gray}
    Accepted for publication at SISAP 2026 (Springer LNCS). This pre-print is not the Version of Record and does not reflect post-acceptance improvements or corrections.
  }\hfil}%
  \let\@evenfoot\@oddfoot
}
\makeatother
\maketitle   % typeset the header of the contribution
\thispagestyle{arxivnote} % Zeigt die Fußnote NUR auf Seite 1 an
\begin{abstract}
The annual SISAP Indexing Challenge benchmarks Approximate Nearest Neighbor Search (ANNS) algorithms under rigorous constraints. This paper presents our submissions for the 2026 edition, addressing both $k$-Nearest Neighbor Graph (kNNG) construction on 1024-dimensional BGE-M3 embeddings (Task 1), and Maximum Inner Product Search (MIPS) on unnormalized Llama-3.2-8B features (Task 2).
Our approach achieved 1st place in Task 2 and 3rd place in Task 1.

To optimize construction speed, we utilize Equi-Voronoi Polytopes (EVP) for efficient quantization, supplemented by targeted reranking strategies to maintain high recall. For MIPS, we transform the asymmetric inner product problem into a Euclidean search space via dimensionality augmentation.
To reduce query latency and optimize memory access, we introduce a 1D presorting mechanism via Fast Linear Assignment Sorting (FLAS) prior to graph construction. This significantly improves spatial locality and cache hit rates during subsequent graph traversal.

Source Code: \url{\repourl}

\end{abstract}

\section{Introduction}
Nearest neighbor search is a cornerstone of modern search systems, recommender engines, and Large Language Models (LLMs). Items are compared using metrics on high-dimensional feature vectors. The computational effort depends on the distance function, dimensionality, and data type. For massive datasets or complex distance functions, exact comparisons become too slow. Therefore, approximate nearest neighbor search (ANNS) algorithms are indispensable.

The annual SISAP Indexing Challenge benchmarks ANNS approaches under strict constraints. The 2026 edition introduces larger datasets and unnormalized feature vectors. Submissions run in Docker containers restricted to 8 vCPUs and 24 GB of RAM on an AMD EPYC 7F72 processor. The challenge comprises three tasks, of which we address the first two. To ensure evaluation consistency, our experiments were conducted on an Intel Xeon 8581C virtual machine configured with identical resource limits and restricted to AVX2 instructions.

\paragraph{\textbf{Task 1}} requires computing an \textit{approximate $k$-Nearest Neighbor Graph} (kNNG) for $k=15$. The dataset contains 6.4 million normalized, 1024-dimensional vectors produced from a BGE-M3 embedding model. Similarity is measured via dot product. The goal is to reduce the total computation time, which includes preprocessing, index building, search, and post-processing. Submissions must achieve at least 0.8 average recall within an 8-hour time limit.

\paragraph{\textbf{Task 2}} evaluates \textit{Maximum Inner Product Search} (MIPS) on LLM-inspired workloads under distribution shifts. The dataset consists of 256,000 unnormalized, 128-dimensional embeddings from a Llama-3.2-8B model. The objective is to find the $k=30$ maximum inner products for a provided set of queries. Unlike Task 1, the evaluation measures only the search phase time. Submissions are limited to 1 hour of runtime and must reach a recall of 0.8. 
\\
\\
The 2025 challenge already featured a $k$-NN graph construction task on 3 million vectors with 384 dimensions. It was won by team \textit{hforest} \cite{Imamura2025}, while refinement-based graph approaches by \textit{BrownCICESE} \cite{Foster2025} proved highly competitive. However, the 2026 tasks raise the bar significantly. Task 1 more than doubles the dataset size to 6.4 million vectors and increases the dimensionality to 1024. This massive scale heavily penalizes slow index construction times. Meanwhile, Task 2 focuses on unnormalized LLM embeddings and a distribution shift between dataset and query vectors.

\section{Related Work}

Exact $k$-Nearest Neighbor Graphs (kNNG) \cite{Paredes2005} connect each vertex to its true closest neighbors but are computationally prohibitive to construct. Therefore, approximate structures like kGraph \cite{Dong2011} are used. Dong et al. introduced the \textit{graph quality} metric to explicitly measure the overlap between these approximated edges and the exact nearest neighbors. However, directly building an index that inherently possesses high graph quality is still slow. Often, a faster strategy is to quickly build a proximity graph, then extract the true top-$k$ neighbors via a local search initialized directly at each target vertex. While the widely used HNSW \cite{Malkov2020} builds rapidly, its hierarchical design prioritizes long-distance navigation edges for out-of-sample queries. When forcing a local exploration search on its bottom layer, the algorithm easily gets trapped in local minima due to missing connectivity guarantees. In contrast, the Dynamic Exploration Graph (DEG) \cite{Hezel2025} is explicitly optimized for this task. It operates on a single, strongly connected layer, ensuring highly efficient local exploration. 
To accelerate graph construction, vector quantization is highly effective; however, traditional methods often require a time-consuming training phase. Connor et al. \cite{Connor2025} propose Equi-Voronoi Polytopes (EVP), a training-free approach mapping high-dimensional vectors to convex polytope vertices. The geometric mapping preserves Spearman similarity correlation and processes vectors independently, ensuring a highly parallelizable conversion. The resulting bit-representations enable rapid distance calculations via Single Instruction, Multiple Data (SIMD) bitwise operations.

\section{Contribution}
In this paper, we address the requirements of the SISAP 2026 Indexing Challenge by optimizing the Dynamic Exploration Graph (DEG) for high-dimensional and LLM-based workloads. Our contributions are summarized as follows:

\begin{itemize}
    \item\textbf{EVP-Accelerated kNNG Construction:} We accelerate high-dimensional proximity graph construction by integrating EVP quantization into the DEG framework. 
    To counteract the resulting precision loss, we design and systematically evaluate hybrid EVP-FP16 configurations to restore target recall.    
        \vspace{0.5em}
    \item\textbf{Dimensionality Augmentation for MIPS:} We apply a lifting technique to embed unnormalized embeddings into a $(d+1)$-dimensional space, transforming the asymmetric inner product maximization into a symmetric distance minimization problem compatible with standard proximity graphs.
        \vspace{0.5em}
    \item\textbf{Cache-Aware Graph Layout:} We introduce a 1D presorting step to arrange vectors by geometric similarity prior to graph construction. Storing similar vectors in contiguous memory locations maximizes spatial locality, boosting cache hit rates and reducing query latency during graph traversal.
\end{itemize}

\section{Task 1 with EVP and DEG}

To minimize the execution time in Task 1, we integrate Equi-Voronoi Polytopes (EVP) quantization \cite{Connor2025} into the Dynamic Exploration Graph (DEG). EVP maps high-dimensional embeddings to ternary values $\{-1, 0, 1\}$, enforcing a constant number of non-zero elements controlled via a hyperparameter. Even asymmetric distance calculations between unquantized FP16 queries and ternary EVP database vectors are highly efficient, requiring only masked additions and subtractions. However, aggressive quantization introduces a precision trade-off by degrading similarity fidelity. Empirical evaluation demonstrates a failure to achieve target recall when relying exclusively on EVP for graph construction and search. The resulting topology requires a hybrid-precision approach and a dedicated reranking phase to recover accuracy while preserving throughput.

\subsection{Evaluated Approaches}
\begin{figure}[h!]
\vspace{-0.5cm}
\centering
\includegraphics[width=\linewidth]{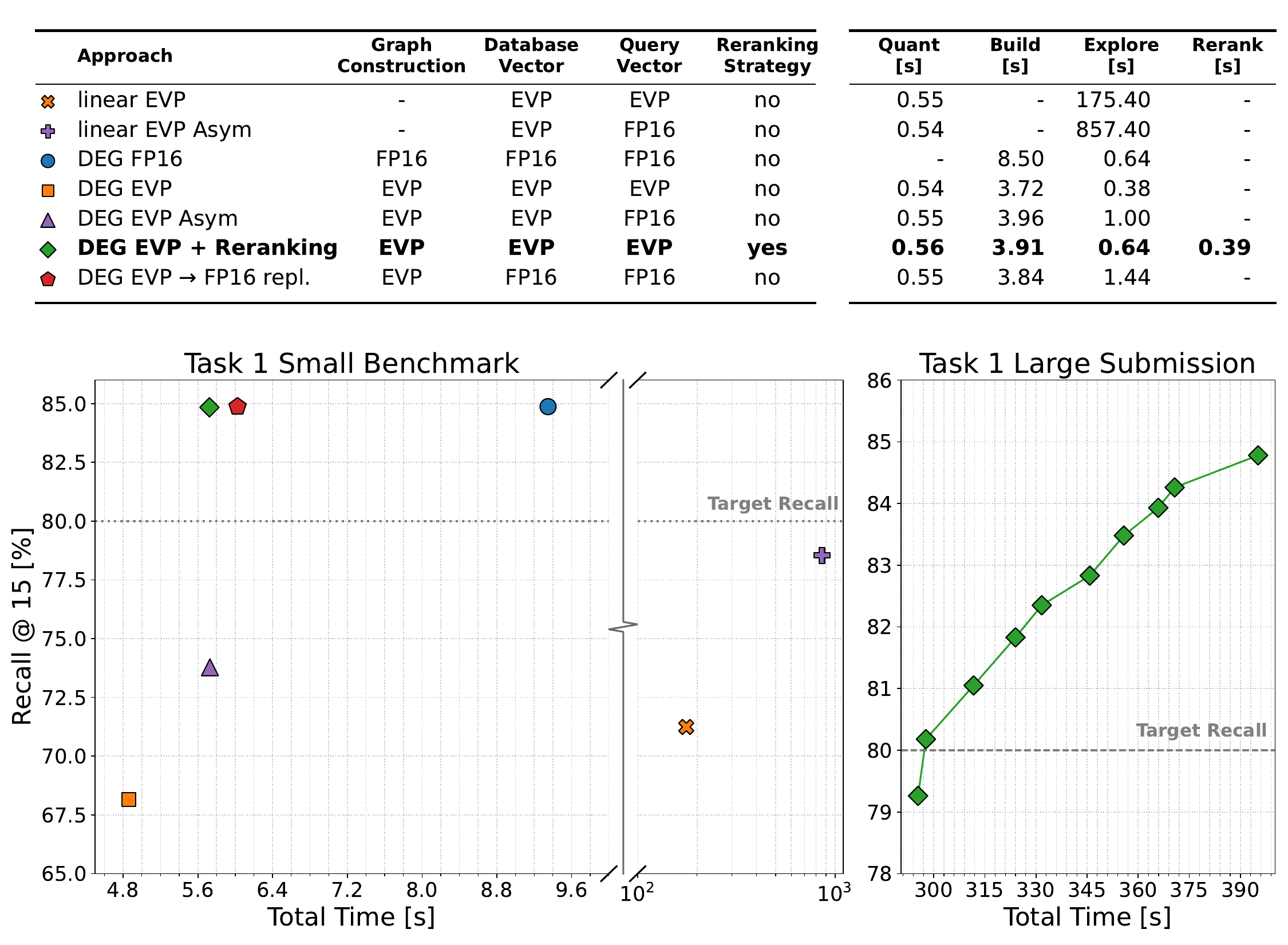}
\caption{Task 1 benchmark results. Top: Algorithmic pipeline configurations and detailed timing breakdowns on the small dataset. Bottom-left: Recall@15 vs.\ total execution time for the small dataset. Bottom-right: Submission Pareto curve for the \textbf{DEG EVP + Reranking} approach across hyperparameter configurations on the large 6.4M vector dataset.}
\label{fig:task1_results}  
\vspace{-0.5cm}
\end{figure}

To evaluate the trade-offs between construction speed and accuracy, we vary vector representation across three pipeline components: \textbf{graph construction}, the \textbf{database vector} format, and the \textbf{query vector}. Five distinct algorithmic configurations are evaluated (top table in Figure \ref{fig:task1_results}):

\begin{itemize}
    \item \textbf{DEG FP16:} Graph constructed, stored, and queried entirely using FP16 vectors. Extracting the 15 nearest neighbors directly from the raw graph edges yields a recall below the 0.8 threshold, while increasing the vertex degree to improve accuracy drastically inflates build times. Therefore, a local exploration search is initiated from each vertex to extract the 15 nearest neighbors without compromising construction speed.
    \vspace{0.5em}
    \item \textbf{DEG EVP:} FP16 vectors are quantized to 2-bit Equi-Voronoi Polytopes (EVP). Both graph construction and the subsequent local exploration search rely entirely on the symmetric EVP metric.
    \vspace{0.5em}
    \item \textbf{DEG EVP Asym:} The initial graph construction matches the symmetric EVP Baseline. During local exploration, original FP16 vectors serve as queries, utilizing an asymmetric distance function against the stored EVP database vectors.
    \vspace{0.5em}
    \item \textbf{DEG EVP + Reranking:} Graph construction operates on EVP vectors. Local exploration retrieves an expanded candidate pool exceeding $k=15$, which is subsequently reranked using exact FP16 features to determine the final nearest neighbors.
    \vspace{0.5em}
    \item \textbf{DEG EVP $\rightarrow$ FP16 repl.:} The index is initially built using EVP vectors. Post-construction, all internal EVP representations are replaced with their original FP16 counterparts. Local exploration then operates exclusively on these float features.
\end{itemize}

%    All    Build   Query
% 1: 1.00x,  1.00x, 1.00x
% 2: 1.08x,  5.54x, 0.80x
% 3: 1.70x, 21.12x, 0.61x

\subsection{Results}

The following results were evaluated on a small provided dataset (about 200,000 vectors) to assess algorithmic efficiency, as shown on the bottom-left of Figure \ref{fig:task1_results}. All graphs share identical construction hyperparameters, and exploration parameters are tuned to reach a recall of 0.85 wherever possible.

Graph construction dominates the total execution time across all experiments. For  \textbf{DEG FP16}, this phase consumes approximately 80\% of the total duration. Utilizing EVP vectors halves the construction time but limits the final recall to 0.68 (\textbf{DEG EVP}). An exhaustive linear search using only EVP bits shows a hard upper bound of 0.71. 
Even applying the asymmetric distance function during linear search raises the upper bound to only 0.78. Both upper bounds fall short of the 0.8 target recall. Furthermore, the asymmetric distance computation is several times slower than the symmetric EVP search. This is true for the linear search as well as the exploration time in (\textbf{DEG EVP Asym}).

Replacing the asymmetric metric in DEG with a subsequent reranking step (\textbf{DEG EVP + Reranking}) yields similar execution times but higher recall rates. For this test, the EVP exploration phase retrieves 50 initial candidates to ensure a high-quality candidate pool for the reranking step. This expanded retrieval pool slightly increases the exploration duration.

Substituting all quantized database vectors with the original FP16 features after graph construction (\textbf{DEG EVP $\rightarrow$ FP16 repl.}) enables standard graph search without a separate reranking phase. The superior precision of the FP16 metric reduces the required graph traversal depth compared to pure EVP. However, the high computational cost of floating-point distance calculations renders this strategy slightly slower than the \textbf{DEG EVP + Reranking} pipeline.
\\
\\
For the large dataset, a hyperparameter search was conducted. The top ten configurations for the \textbf{DEG EVP + Reranking} approach are identified and shown in Figure \ref{fig:task1_results} on the bottom-right. Achieving the 0.8 target recall requires a total execution time of 297 seconds. This duration comprises 52\% graph construction, 5\% EVP quantization, 34\% exploration and reranking, and the remaining fraction for disk loading.

\section{Task 2 with Dimension Expansion and FLAS}
For Task 2, we adapted our graph strategy from Task 1. However, unlike the first task, it evaluates \textit{Maximum Inner Product Search} (MIPS) on unnormalized Llama-3.2-8B embeddings, where varying vector norms distort traditional proximity spaces. Preliminary baseline experiments using an exhaustive linear search with EVP yielded a recall of only 0.08, demonstrating how direction-only quantization strips the magnitude information required for MIPS.

To address these geometric constraints and maximize search throughput, we investigate two independent architectural optimizations:

\begin{itemize}
    \item \textbf{Dimension Augmentation:} A lifting technique inspired by the ALSH framework \cite{Shrivastava2014} maps unnormalized vectors into a higher-dimensional space. This transformation stabilizes vector norms, enabling evaluation via both Inner Product and Euclidean distance metrics.
    \vspace{0.5em}
    \item \textbf{Cache-Aware Layout Optimization:} To maximize hardware throughput during the search phase, we apply a spatial reordering strategy via the FLAS algorithm \cite{Barthel2023} prior to graph construction. This optimizes cache line utilization during traversal and is evaluated on both the original and augmented feature spaces.
\end{itemize}

\subsection{Transforming Inner Products into Euclidean Distances}
Maximum Inner Product Search (MIPS) can be reduced to Euclidean Nearest Neighbor Search (NNS) via dimensionality augmentation. For each $d$-dimensional vector $\mathbf{v}$, we define $\tilde{\mathbf{v}} = [\mathbf{v}^\top, \sqrt{M^2 - \|\mathbf{v}\|^2}]^\top$, where $M \ge \max \|\mathbf{v}\|$. 
The query is augmented as $\tilde{\mathbf{q}} = [\mathbf{q}^\top, 0]^\top$.
The squared Euclidean distance satisfies:
\begin{equation}
\|\tilde{\mathbf{q}} - \tilde{\mathbf{v}}\|^2 = \|\mathbf{q}\|^2 + M^2 - 2\mathbf{q}^\top\mathbf{v}
\end{equation}

\begin{figure}[h!]
\centering
\includegraphics[width=1\linewidth]{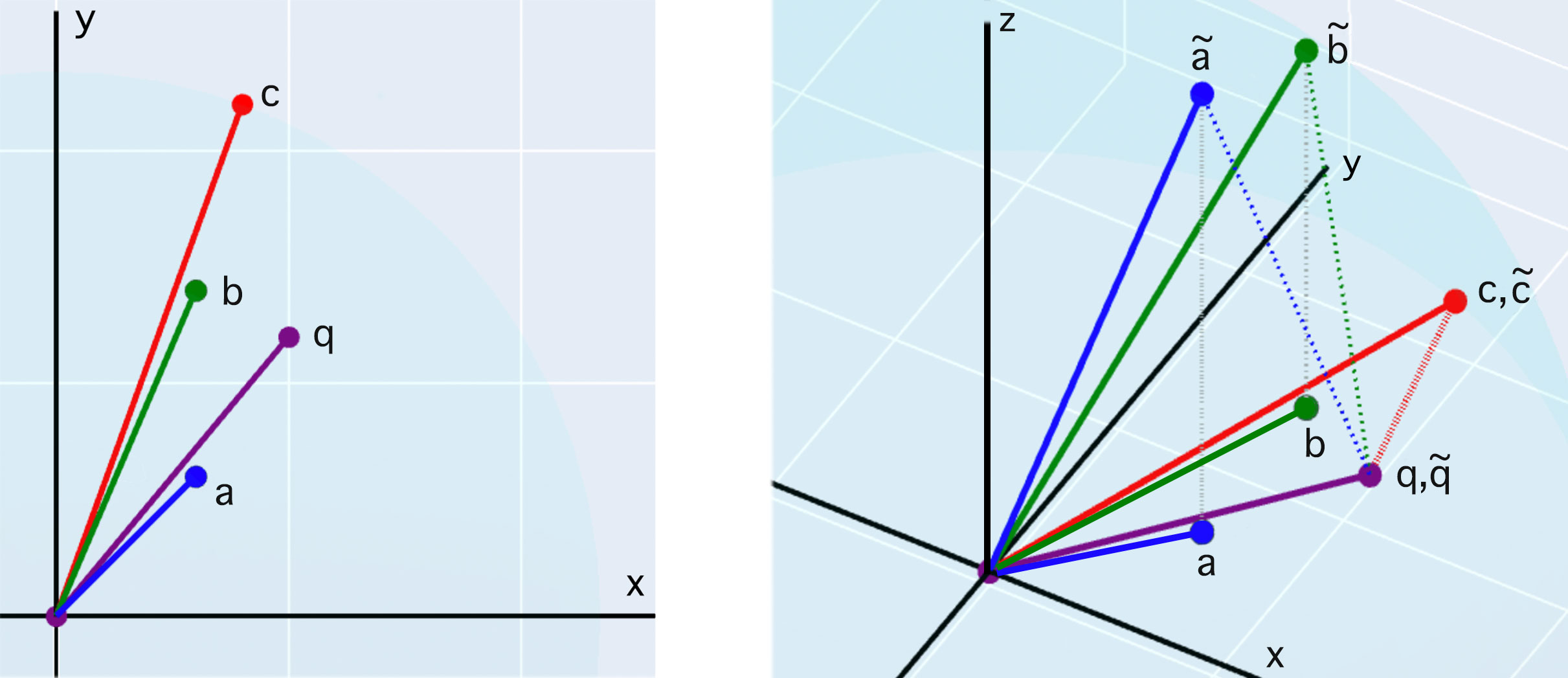}
\caption{Geometric embedding of MIPS. Left: In 2D, $\mathbf{c}$ is the vector maximizing the inner product with $\mathbf{q}$. Right: The lifted vectors $\tilde{\mathbf{a}}$, $\tilde{\mathbf{b}}$, and $\tilde{\mathbf{c}}$ reside on a 3D sphere, where minimizing the Euclidean distance to the augmented query $\tilde{\mathbf{q}}$ recovers the maximum inner product.}\label{fig:dim_expansion}
%\vspace{-0.5cm}
\end{figure}

Since $\|\mathbf{q}\|^2$ and $M^2$ are constant for a fixed query, minimizing $\|\tilde{\mathbf{q}} - \tilde{\mathbf{v}}\|^2$ is equivalent to maximizing $\mathbf{q}^\top\mathbf{v}$. Geometrically, this maps vectors onto a $(d+1)$-dimensional hypersphere of radius $M$, rendering the problem directly compatible with standard Euclidean proximity graphs. Figure~\ref{fig:dim_expansion} shows an example of this MIPS reduction in 2D.

\subsection{Cache-Aware Graph Construction via 1D Presorting}

While the Dynamic Exploration Graph efficiently handles evolving datasets, its default construction phase processes vectors in an arbitrary insertion order. To optimize the index for modern memory hierarchies in static workloads, we introduce a presorting step prior to vertex insertion, utilizing the Fast Linear Assignment Sorting (FLAS)\footnote{\url{https://github.com/Visual-Computing/LAS_FLAS/}} algorithm \cite{Barthel2023} to sort all vectors one-dimensionally by their similarity.

This strategy maps geometrically similar vectors onto contiguous physical memory addresses. Constructing the DEG from this presorted sequence ensures local neighborhoods align with consecutive cache lines. Consequently, the resulting spatial locality improves CPU cache hit rates and mitigates high-cost main memory fetches, significantly reducing overall query latency.

\subsection{Results and Evaluation} \label{subsec:task2_results}

\begin{figure}[h!]
\centering
\includegraphics[width=\textwidth]{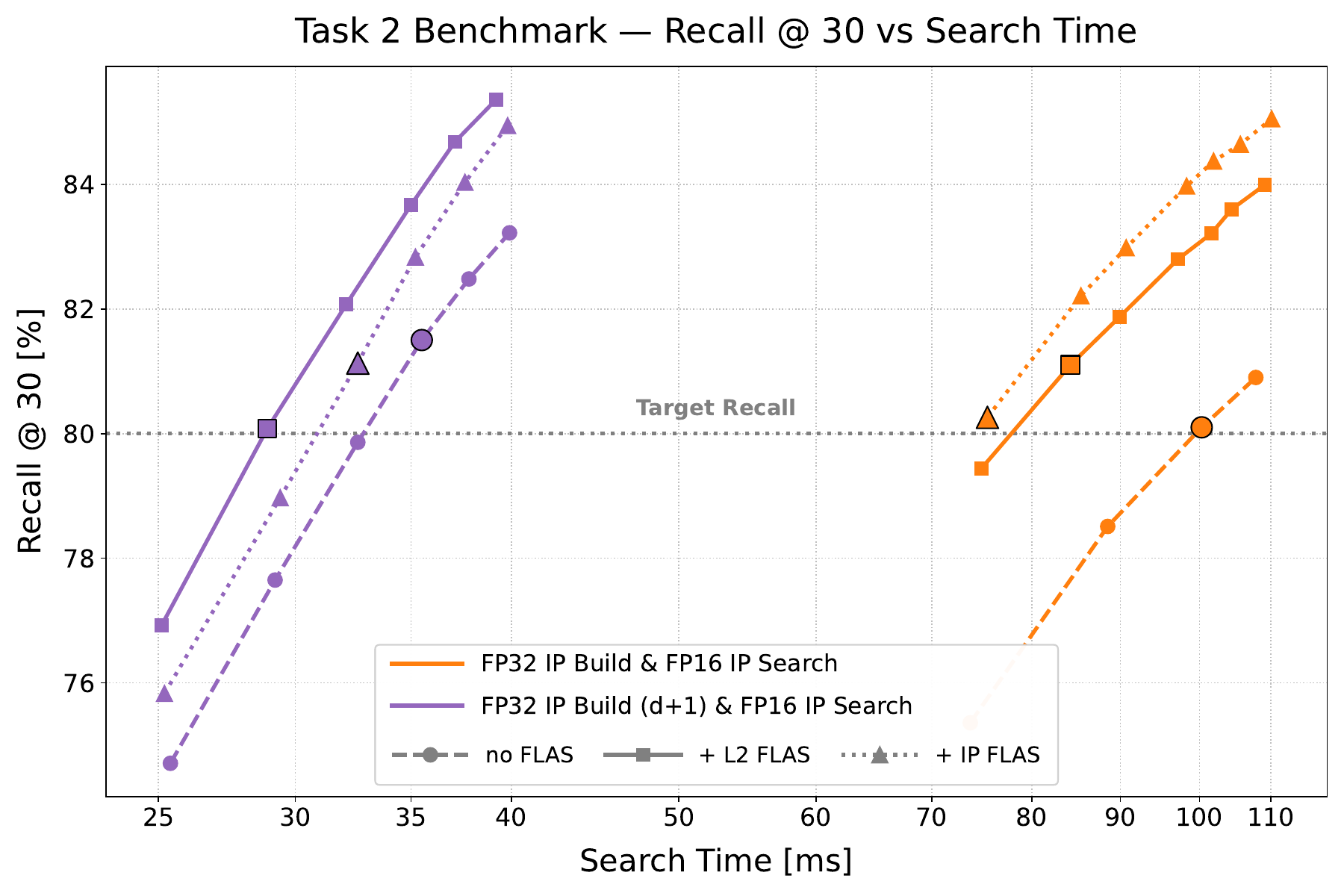}
\caption{Task 2 query latency versus recall, comparing the baseline MIPS against FLAS sorting types on the right and  dimension expansion + FLAS approaches on the left.} 
\label{fig:task2_results}  
\end{figure}

The evaluation for Task 2 focuses on the trade-off between search latency and recall. Figure~\ref{fig:task2_results} illustrates the performance of the baseline graph against our layout and dimensionality optimization strategies.

The baseline Dynamic Exploration Graph, constructed in the original unnormalized FP32 space and searched via the Inner Product metric using FP16 features, requires approximately 100~ms to complete all queries due to the distorted geometric space. Integrating the FLAS sorting step prior to graph construction reduces query latency to 80~ms, achieving a 20\% performance gain solely through enhanced spatial locality. Within this unnormalized space, configuring FLAS to utilize the Inner Product metric yields superior results compared to an $L_2$ configuration, as aligning the sorting metric with the underlying search space preserves local neighborhood structures.
Applying dimension augmentation without presorting slashes search time to 33~ms, a 67\% improvement over the baseline. This reduction demonstrates how lifting embeddings into the $(d+1)$-dimensional space stabilizes vector norms and resolves the geometric discrepancies of unnormalized MIPS. Crucially, within this augmented space, the optimal metric for FLAS shifts to $L_2$. Because the wide dynamic range of Inner Product values introduces severe quantization inaccuracies during the FLAS sorting phase, employing the bounded $L_2$ metric post-extension mitigates these precision errors.

Compounding both methodologies delivers the highest performance. Combining dimension extension with the $L_2$-configured FLAS reordering reduces query latency to 29~ms, achieving an additional 12\% reduction over the unsorted, lifted approach.

\section{Conclusion}

In this paper, we presented our contributions to the SISAP 2026 Indexing Challenge by optimizing the Dynamic Exploration Graph (DEG) for high-di\-men\-sional, distribution-shifted LLM workloads. For Task 1 (normalized embeddings), we utilized Equi-Voronoi Polytopes (EVP) quantization to accelerate graph construction. Supplemented by an FP16 reranking step, this approach reduced total execution time on the small dataset from 9.35~s to 5.72~s, a 39\% performance improvement, while maintaining target recall. For Task 2 (unnormalized MIPS), we resolved geometric discrepancies by lifting the feature dimensions, cutting baseline query latency from 100~ms to 33~ms. Compounding this transformation with a spatial data reordering step via the FLAS algorithm further reduced latency to 29~ms by maximizing CPU cache utilization. Our results demonstrate the effectiveness of decoupling geometric transformations and cache-aware memory layouts from the baseline graph topology to optimize performance trade-offs.

\subsubsection*{Disclosure of Interests.}
The authors have no competing interests to declare that are relevant to the content of this article.
\newpage

% ---- Bibliography ----
%
% BibTeX users should specify bibliography style 'splncs04'.
% References will then be sorted and formatted in the correct style.
%
\bibliographystyle{splncs04}
\bibliography{references}
\end{document}